# Tunable Unidirectional Nonlinear Emission from Transition-Metal-Dichalcogenide Metasurfaces

Mudassar Nauman[a,b], Jingshi Yan[b], Domenico de Ceglia[c], Mohsen Rahmani[b,d], Khosro Zangeneh Kamali[b], Costantino De Angelis[e], Andrey E. Miroshnichenko[e], Yuerui Lu[a] and Dragomir N. Neshev[b]

[a] *Research School of Electrical, Energy and Materials Engineering, College of Engineering and Computer Science, Australian National University, Canberra ACT, 2601, Australia*
[b] *ARC Centre of Excellence for Transformative Meta-Optical Systems, Department of Electronic Materials Engineering, Research School of Physics, The Australian National University, Canberra ACT, 2601, Australia*
[c] *Department of Information Engineering, University of Padova, Via G. Gradenigo, 6/B, Padova, Italy*
[d] *Advanced Optics and Photonics Laboratory, Department of Engineering, School of Science and Technology, Nottingham Trent University, Nottingham, NG11 8NS, UK*
[e] *Department of Information Engineering, University of Brescia, Via Branze 38, 25123 Brescia, Italy*
[f] *School of Engineering and Information Technology, University of New South Wales, Canberra ACT, 2600, Australia*

dragomir.neshev@anu.edu.au  yuerui.lu@anu.edu.au, andrey.miroshnichenko@unsw.edu.au, domenico.deceglia@unipd.it

**Abstract:**

Nonlinear light sources are central to a myriad of applications, driving a quest for their miniaturisation down to the nanoscale. In this quest, nonlinear metasurfaces hold a great promise, as they enhance nonlinear effects through their resonant photonic environment and high refractive index, such as in high-index dielectric metasurfaces. However, despite the sub-diffractive operation of dielectric metasurfaces at the fundamental wave, this condition is not fulfilled for the nonlinearly generated harmonic waves, thereby all nonlinear metasurfaces to date emit multiple diffractive beams. Here, we demonstrate the enhanced single-beam second- and third-harmonic generation in a metasurface of crystalline transition-metal-dichalcogenide material, offering the highest refractive index. We show that the interplay between the resonances of the metasurface allows for tuning of the unidirectional second-harmonic radiation in forward or backward direction, not possible in any bulk nonlinear crystal. Our results open new opportunities for metasurface-based nonlinear light-sources, including nonlinear mirrors and entangled-photon generation.

**Introduction**

Nonlinear optical phenomena in nanostructured materials have recently attracted much attention due to their wide spectrum of possible applications[1,2] ranging from sensing to novel light sources. In these studies, the enhancement of the nonlinear processes at the nanoscale has been a key enabler. The two main paths in this development have been the use of high-quality factor resonant photonic structures and highly nonlinear materials, including high refractive-index materials following the so-called Miller's rule. The earlier approaches were based on the use of plasmonic nanoantennas and metasurfaces[3,4,5,6] as well as multiple-quantum-well (MQW) metasurfaces[7,8,9]. However, the high dissipative losses of plasmonic materials and the mid-infrared operation wavelengths of MQWs metasurfaces limit their broad applicability. More recently, high-index dielectric nanostructures have opened a new paradigm for nonlinear meta-optics[10,11]. The excitation of Mie-type resonances in such nanostructures plays a vital role in enhancing the light-matter interactions, such as nonlinear harmonic generation[12,13,14]. The use of high-index dielectric materials, such as semiconductors, have simultaneously provided a strong intrinsic nonlinear susceptibility and high-quality factor resonances. Several pioneering works prompted the use of nanostructures from materials with established fabrication to enhance the process of third harmonic generation (THG). Among them, Si[15,16,17] and Ge[18] show the high refractive index in the near-infrared spectral range and exhibit strong nonlinear response. However, Si and Ge are centrosymmetric materials and thereby inhibit the second-order nonlinear effects[7].

Conversely, zinc-blende III-V semiconductors, such as GaAs and AlGaAs have received a great deal of interest for nonlinear metasurfaces due to their high-refractive indices and strong quadratic susceptibilities[19,20,21,22,23,24,25]. However, zinc-blende III-V nanostructures are difficult to fabricate on low-index transparent substrates and exhibit a peculiar nonlinear susceptibility with only off-diagonal non-zero $\chi^{(2)}$ tensor components[20]. As a consequence, in (100) AlGaAs metasurface, the second-harmonic generation (SHG) is inhibited at the normal direction and the harmonic emission is directed into the diffraction orders[26,27]. This nonlinear diffraction phenomenon limits the possible applications of metasurfaces as nonlinear light sources, where single beam of emission is required. While metasurfaces made of inversion-symmetry-broken nanoresonators[28,29] and individual nanoantennas[30,31] have been proposed to direct the emission at normal direction, to date they operate in the diffractive regime[12,13,14], with multiple diffractive orders being emitted at the harmonic waves[27,32]. The ability to operate in the zero$^{\text{th}}$-order SHG and THG, as well as to be able to tune the directionality of such unidirectional nonlinear emission from forward to backward direction remains elusive.

Here, we demonstrate, for the first time to our knowledge, the significant enhancement of SHG and THG in a transition-metal-dichalcogenide (TMDC) MoS$_2$ metasurface in the sub-diffractive, single-beam nonlinear emission regime. Our metasurfaces consist of nanoresonators with the highest refractive index >4, in the visible range, prompting to strong material nonlinearity. The high refractive index further allows for optical resonances in smaller-size nanoresonators, thereby for densely packed metasurfaces. In particular, we design our metasurface to exhibit multipolar resonances at the second-harmonic wavelengths and employ the interference of these resonance to enable the *tuning of the unidirectional emission of the second-harmonic light in forward or backward direction*, controlled either by the excitation wavelength or by the incident polarisation. In all operation regimes, our metasurface remains sub-diffractive for the SHG and THG wavelengths. Finally, our TMDC metasurface can be fabricated on any transparent substrate due to the strong van der Waals (vdW) stiction forces of the layered MoS$_2$. As such, our TMDC metasurfaces provide a versatile tool to tune all harmonic emissions and open new applications of nonlinear metasurfaces for advanced light sources[33] or nonlinear mirrors[34].

**Results**

*Linear response.* TMDCs are vdW materials of high refractive index with relatively large, indirect bandgap in the visible spectrum. They have attracted significant attention for high-performance photonic devices, such as miniaturised lenses[35,36] and gratings[37]. More recently, TMDCs have sparked an interest for novel dielectric metasurfaces with high refractive index[38,39,40,41,42,43,44] due to the opportunity to fabricate high purity single crystalline films on any transparent substrate. The TMDC materials, such as MoS$_2$ further exhibit low optical losses in the visible and near infrared spectral regions. The refractive index of the MoS$_2$ is over 4.5 (Fig. S1 of Supplementary Information) in the region 700-800 nm, which is significantly higher than in other nonlinear metasurfaces, e.g. those based on Si, GaAs or AlGaAs, and promotes strong nonlinear response.

Here we fabricate a TMDC-based MoS$_2$ metasurface, which supports optically induced resonant modes in the visible spectrum, where MoS$_2$ has a negligible absorption. According to Mie theory[10] these resonances depend upon the refractive index and the geometric parameters of the individual meta-atoms[45]. We start with exfoliating high-quality multilayer MoS$_2$ flakes on a sapphire substrate. The flake is then patterned using electron beam lithography (EBL) and reactive-ion etching (see Methods). The design of our nonlinear metasurface is illustrated in Fig. 1a. It consists of a periodic square array of truncated cone meta-atoms, placed on low refractive index sapphire ($Al_2O_3$) substrate (~500 µm thick). A scanning electron microscopy (SEM) image of a typical metasurface is shown in Fig. 1b. Metasurfaces of three different geometric dimensions have been fabricated (A-C), exhibiting resonances from the visible to the near-infrared spectral range.

The linear transmission spectra of the fabricated metasurfaces (A, B and C having same height ≅ 150 nm but different radii 100 nm, 110 nm, 120 nm, respectively) were measured in a custom-built white light transmission spectroscopy setup and are shown in Fig. 1c. The white light is focused onto a sample and collected by 20× microscope objective lenses (NA=0.4). The spectra are recorded by two different spectrometers for the visible and infrared part of the spectrum, respectively. All three metasurfaces show distinct Mie-type resonances in the spectral range of 700-800 nm. We have also performed numerical simulations (Lumerical FDTD) to systematically study the resonant modes of the fabricated $MoS_2$ metasurfaces (Fig. S1 of Supplementary Information). The induced resonances are associated with transmittance dips and marked by arrows in Fig. 1c. By changing the aspect ratio (periodicity is changed together with the diameter of meta-atoms, whilst keeping the height unchanged) of the $MoS_2$ truncated cones (for metasurfaces A to C) we can control the spectral position of the resonances from visible to infrared wavelengths[46]. All metasurfaces are fully transparent at longer wavelength (>900 nm) (Fig. S2 of Supplementary Information).

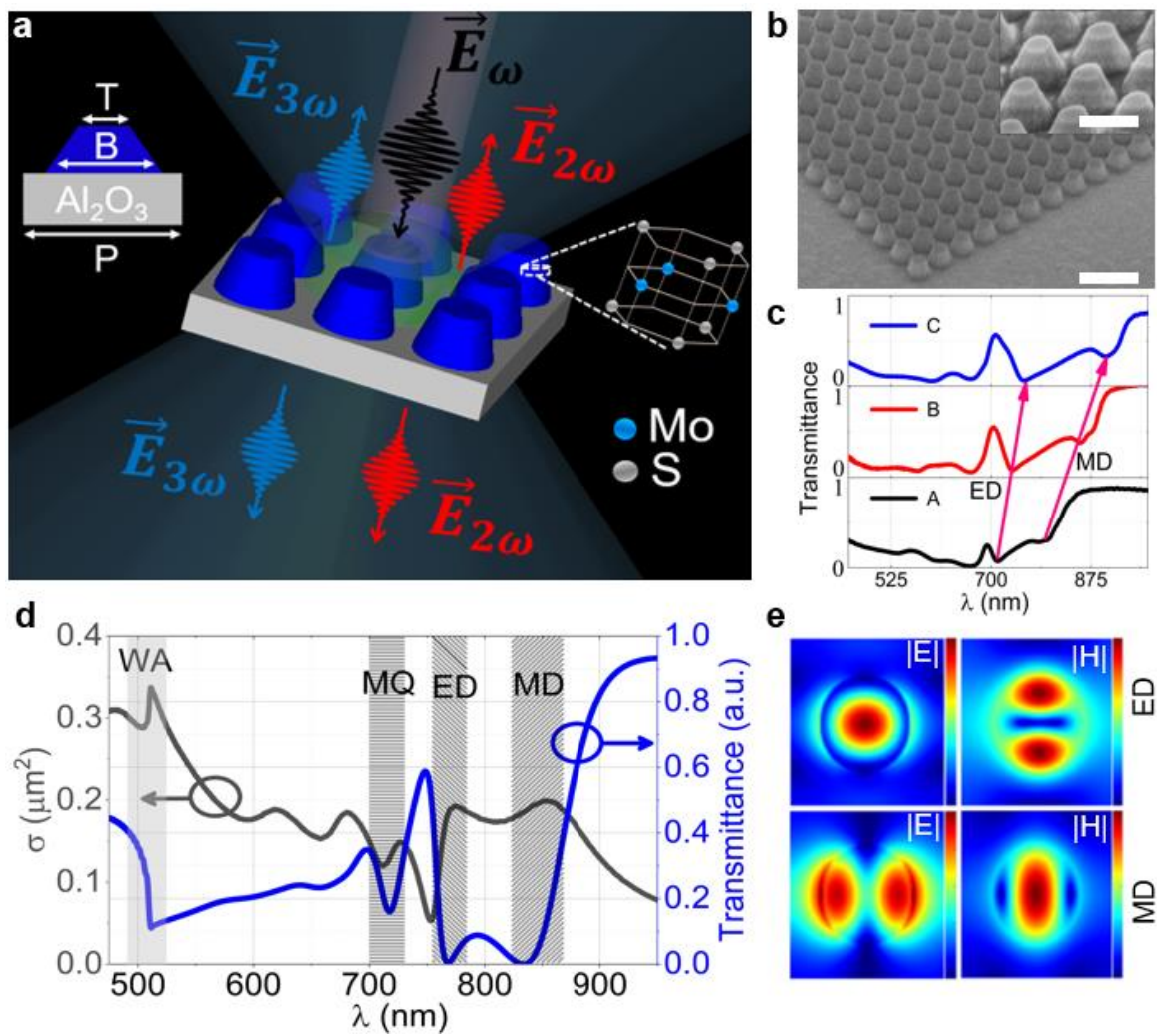

**Fig. 1**. **Nonlinear TMDC metasurfaces.** (a) Schematic of $MoS_2$ truncated cone metasurface. For metasurface C, $T$ = 150 nm, $B$ = 240 nm, $P$ = 300 nm and height = 150 nm, as shown in the top left corner. The bottom right inset shows the lattice structure of the 2H-stacked $MoS_2$. (b) The SEM image of the fabricated metasurface C with a scale bar of 2 $\mu m$ while inset scale bar is 500 nm. (c) The measured transmission spectrum of three metasurfaces A, B, and C. For clarity spectra are vertically displaced by $T$ = 1 for each metasurface. Arrows show the change in the spectral position of electric and magnetic modes as a function of the aspect ratio of truncated cone meta-atoms. (d) Calculated cartesian multipoles expansion of metasurface C, where $\sigma$ represents the effective scattering cross-section. (e) Normalised electric and magnetic field profiles at the resonant wavelengths: 770 nm and 840 nm, respectively.

The nature of the resonances can be further verified by examining the cartesian multipole expansion of the near fields of metasurface C. The linear scattering is illustrated in Fig. 1d by the effective scattering cross-section, σ (black curve) and the corresponding transmittance through the metasurface. A qualitative agreement is observed between the measured and simulated results (Fig. S1 of Supplementary Information). This agreement further allows us to identify the nature of the resonances by exploring the symmetry of the electric and magnetic fields. These are plotted in Fig. 1e, showing the distributions of electric and magnetic fields in a horizontal cut-plane through the centre of the truncated cone meta-atom. The resonance mode at shorter wavelengths has a maximum of the electric field in the centre of the resonator and correspond to an electric dipole (ED) mode. The longer wavelength resonance mode has, in contrast, maximum of the magnetic field in the centre of the resonator and is linked to a magnetic dipole (MD) mode. Additionally, all metasurfaces support higher-order Mie-type resonances, for wavelengths below 700 nm. These higher-order resonances, however, fall in the region where the $MoS_2$ possess excitonic resonances and higher absorption, see also the $MoS_2$ dispersion in Fig. S1 of Supplementary Information. Thereby, experimentally these high-order Mie resonances are difficult to identify and are not marked in the experimental spectrum, Fig. 1c. Nevertheless, for metasurface C the magnetic quadrupole (MQ) mode can be seen as a small shoulder around 715 nm, in Fig. 1c. The total extinction cross section obtained by multipolar decomposition in Fig. 1d shows dominant ED, MD, and MQ contributions of the resonances.

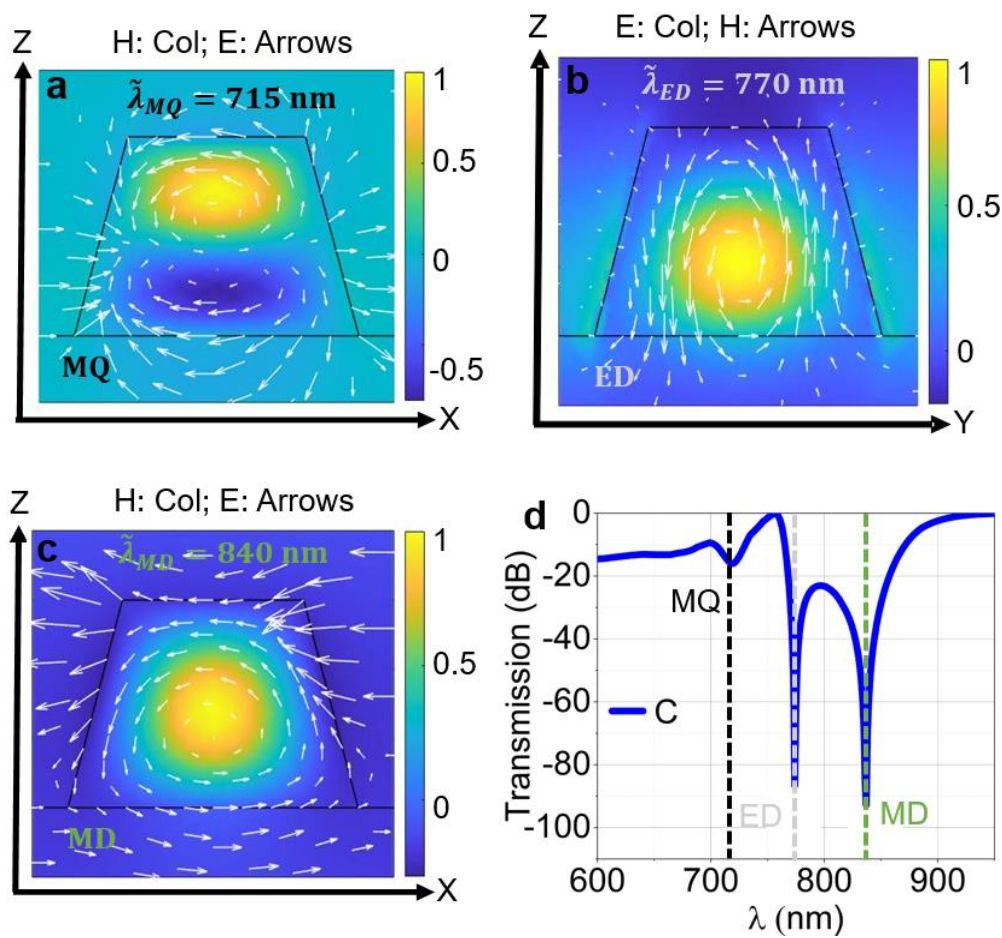

**Fig. 2. Quasi-normal modes for metasurface C.** (a) Eigenfields (electric and magnetic field distribution) of the eigenwavelength ($\tilde{\lambda}_{MQ}$) is reminiscent of MQ mode , E: shown as arrows and H: shown as colour (b) Eigenfields (electric and magnetic field distribution) of the eigenwavelength ($\tilde{\lambda}_{ED}$) is reminiscent of ED mode , E: shown as arrows and H: shown as colour (c) Eigenfields (electric and magnetic field distribution) of the eigenwavelength ($\tilde{\lambda}_{MD}$) is reminiscent of MD mode , E: shown as arrows and H: shown as colour (d) Simulated transmission spectra calculated for plane-wave excitation at normal incidence for metasurface C. Very close to the real eigenwavelengths (shown by vertical dashed lines, following the colour coding scheme of the eigenwavelengths (shown in Fig.2a-c)) we have three transmission dips. The Fano-like nature of these three transmission dips (around 715 nm, 770 nm and 840 nm) is clear indication of the excitation of the three QNMs. As the eigenfields of these QNMs are reminiscent of the typical field distribution of MQ, ED and MD modes and for this reason we have labelled the three QNMs as "MQ", "ED" and "MD" at the bottom inset of Fig. 2a-c, respectively.

However, all resonant modes are generally of mixed nature and to better identify their character we further performed quasi-normal mode (QNM) theory simulations[47] for metasurface C. The QNM representation allows us to quantify the coupling of the modes to free space and determine the role and weight of the individual modes into the harmonic radiation. To retrieve the characteristic wavelength and the lifetime of the QNM we consider the metasurface as an open-cavity system that supports eigenmodes and solve this complex eigenvalue (i.e., source-free) problem with finite-element numerical simulations (COMSOL). The eigenvalues of the problem are the QNMs complex eigenfrequencies $\widetilde{\omega}_m = Re(\widetilde{\omega}_m) + iIm(\widetilde{\omega}_m)$, the real part being associated with the characteristic wavelengths $\tilde{\lambda}_m = 2\pi c/Re(\widetilde{\omega}_m)$ of the modes and the imaginary part associated with the modes lifetimes $\tilde{\tau}_m = 1/Im(\widetilde{\omega}_m)$. The knowledge of the eigenvalues $\widetilde{\omega}_m$ and the normalised eigenfields $\widetilde{\boldsymbol{E}}_m$ associated with QNMs, allows to decompose the radiation of the metasurface at these wavelengths, such as the SHG, as follows:

$$\boldsymbol{E}_{rad} = \sum_m \alpha_m \widetilde{\boldsymbol{E}}_m, \quad (1)$$

where the weighting factors, $\alpha_m$, are frequency-dependent functions proportional to the overlap integrals of the eigenmodes with the background field $\boldsymbol{E}_b$ generated by the source in the absence of the resonators (in our case, the system composed simply by an interface between air and the sapphire substrate). The eigenfields are normalised as in Ref.[48]. We further adopt the following formulation $\alpha_m = -\frac{\widetilde{\omega}_m}{\widetilde{\omega}_m - \omega_{rad}} \int_V \widetilde{\mathbf{E}}_m \cdot \mathbf{E}_b dV$, where $\omega_{rad}$ is the angular frequency of the radiation. Our analysis shows that the $MoS_2$ metasurface supports three QNMs with strong field confinement in the range of wavelengths between 700 nm and 900 nm. The eigenfields distributions associated with these eigenmodes are reminiscent of the field distributions of MD, ED and MQ-type Mie modes, as shown in Fig. 2a-c. In particular, we find an eigenmode at $\tilde{\lambda}_{MD} = 840$ nm with an MD-like eigenfield and lifetime $\tau_{MD} = 6.7$ fs. The second eigenmode at $\tilde{\lambda}_{ED} = 770$ nm has an ED-like eigenfield and lifetime $\tau_{ED} = 21$ fs. The third eigenmode at $\tilde{\lambda}_{MQ} = 715$ nm has an MQ-like eigenfield and lifetime $\tau_{MQ} = 17$ fs. In Fig. 2d, we show, for the same metasurface, the transmission spectrum for plane-wave excitation at normal incidence. The spectrum displays three transmission dips very close to the real eigen-wavelengths of the three QNMs. The signatures of the excitation of the three QNMs are the Fano-like resonances labelled as MD, ED and MQ. Such labels for the QNMs are only introduced to underscore the similarity of the eigenfields to the typical field distributions of multipolar modes. The ED, MD and MQ-like QNMs supported by the metasurface play a crucial role not only in the linear response (Fig. 2d), but also in enhancing the second-order nonlinear response obtained when the pump is tuned at double those eigen-wavelengths, as discussed in the next section titled "*SHG in TMDC metasurfaces*".

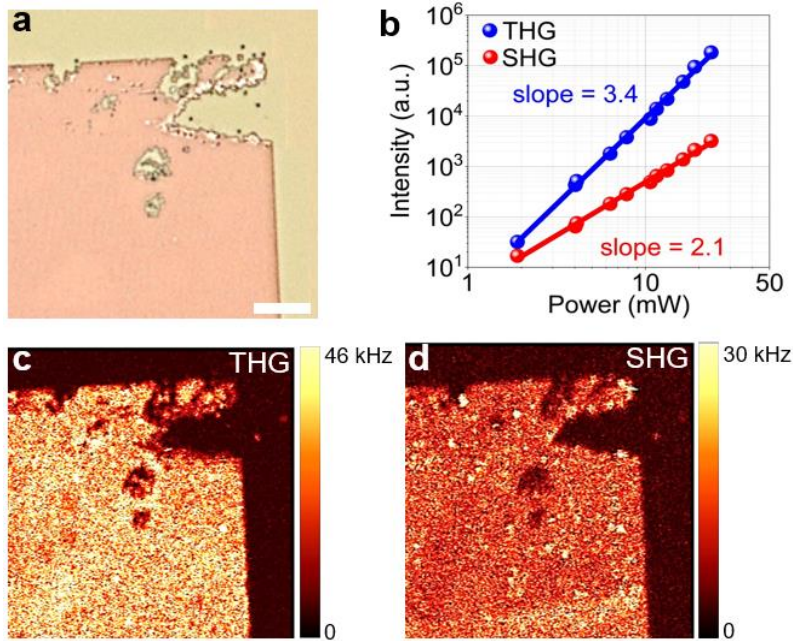

**Fig. 3. Nonlinear microscopy of TMDC metasurfaces.** (a) Optical microscopic image of $MoS_2$ truncated cone metasurface C, scale bar is $10\mu m$. (b) Power dependent nonlinear SH and TH signals from the pump laser at 1550 nm. (c, d) Third- and second-order nonlinear spatial-mapping of the metasurface C, respectively.

*Nonlinear response of TMDC metasurfaces.* Next, we test the nonlinear frequency conversion from the TMDC metasurface. The optical microscope image of metasurface C is shown in Fig. 3a. We specifically chose the top-right corner of the metasurface, which shows a fabrication defect. Such a defect, however, allows for the spatial identification of the nonlinear emission from the metasurface. Next, we illuminate the metasurface by a 10 ps laser pulses from a fibre laser and amplifier (Pritel, a repetition rate of 20 MHz), operating at telecommunication wavelength of 1550 nm. The pump is focused onto the sample by a 100× microscope objective (NA=0.7), and the nonlinear signal is collected in transmission. The measured power for the SH and TH signals shows quadratic and cubic dependence on the incident power, respectively, as shown in Fig. 3b. No saturation of the emitted SH and TH signals is observed in Fig. 3b when the pump average power is increased up to 24 mW, corresponding to a peak pump power of 120 W (peak intensity of ~0.95 GW/cm$^2$). The sample is embedded in a confocal scanning microscope (WITec) that allows for performing nonlinear imaging microscopy by scanning the sample stage. The THG and SHG nonlinear images of the MoS$_2$ metasurface C are shown in Fig. 3c and 3d, respectively. The obtained THG (517 nm) and SHG (775 nm) signals are at wavelengths where the metasurface exhibits the first-order Wood-anomaly in the substrate and the ED-like QNM, respectively. We note that the third harmonic is significantly enhanced by the metasurface, well above the background signal from the substrate. The second harmonic response of the bulk MoS$_2$ depends upon the number of layers in the MoS$_2$ flake. A flake with odd number of layers is non-centrosymmetric and it belongs to $\boldsymbol{D_{3h}^1}$ space group[49,50]. An MoS$_2$ flake with even number of layers is centrosymmetric and belongs to $\boldsymbol{D_{3d}^3}$ space group[49] thereby it does not exhibit second-order nonlinaerity[49,51]. However, the metasurface appears to result in a moderate amount of SHG, again well above the background of the substrate. While recent works[41] suggest that the SHG from TMDC nanoresonators has a bulk response, such possibility is unlikely and requires mode detailed analyses, as shown below.

It is worth noting that to provide sub-diffractive nonlinear response, the periodicity of the metasurface should be, $P < \lambda/n$, where $P$ is the metasurface periodicity, $n$ is the refractive index of the surrounding environment, being either air or sapphire substrate in our case, and $\lambda$ is the wavelength of the harmonic wave (THG or SHG). This condition stems from the equation for the diffraction angle $\theta_{diff}$,

$$\theta_{diff} = \sin^{-1}\frac{\lambda_{SH}}{P \times n}. \qquad (2)$$

As compared to other lower-index semiconductor metasurfaces the use of the high-index TMDCs allows us to design highly subwavelength resonators and therefore to construct nonlinear metasurfaces with sub-diffractive properties at the harmonic waves.

*THG in TMDC metasurfaces.* First, we examine the enhancement of the THG process in TMDC metasurface. For this purpose, we use a tunable femtosecond laser (Chameleon Ultra II and OPO, pulse width of ~200 fs) in a home-build microscopy setup. We performed wavelength dependent nonlinear measurements by focusing the laser beam onto the sample using 20× microscope objective (NA=0.4). We then collect the third harmonic in the forward direction by a second objective (100×, NA=0.7). To filter out the transmitted and reflected pump wave, we employed a short-pass filter at 800 nm. Two waveplates are used to control the polarisation of the pump beam. Figure 4a depicts the spectra of the THG from metasurface C for different pump wavelengths. The black curve represents a spline through all maxima. Two distinct regions could be observed in THG: below and above the Wood anomaly (WA)[52] marked with grey shading in Fig. 4a. Below the WA, our metasurface operates in the diffractive regime, where a large amount of the THG is emitted into the diffraction orders and not collected by the microscope objective and the detector. At the WA, the third harmonic is emitted in the plane of the metasurface and no light is detected. For wavelengths above the WA, the metasurface is sub-diffractive, and the THG is emitted into the zeroth order only, which shows a pronounced enhancement at 517 nm. Measurements above 530 nm (>1590 nm pump wavelength) are unfortunately not accessible by our pump laser. In our experiments, the highest third harmonic power is obtained at wavelength of 1550 nm, resulting in THG efficiency of $1.01 \times 10^{-9}$ at peak power of 4.3 kW, beam full width at half maximum (FWHM) of 4.6 µm, see Fig. S3 of Supplementary Information. While the absolute efficiency is lower than the previous works on THG in single silicon nanostructures, the current value is obtained at five times lower intensity. Considering the quadratic dependence of the efficiency on the intensity and the lack of resonances at the fundamental wavelengths, the obtained value is impressive. In Fig. 4a, the THG intensity is normalised to the cubic power of the incident beam intensity, to avoid the small variations of the incident power, while tuning the OPO wavelength.

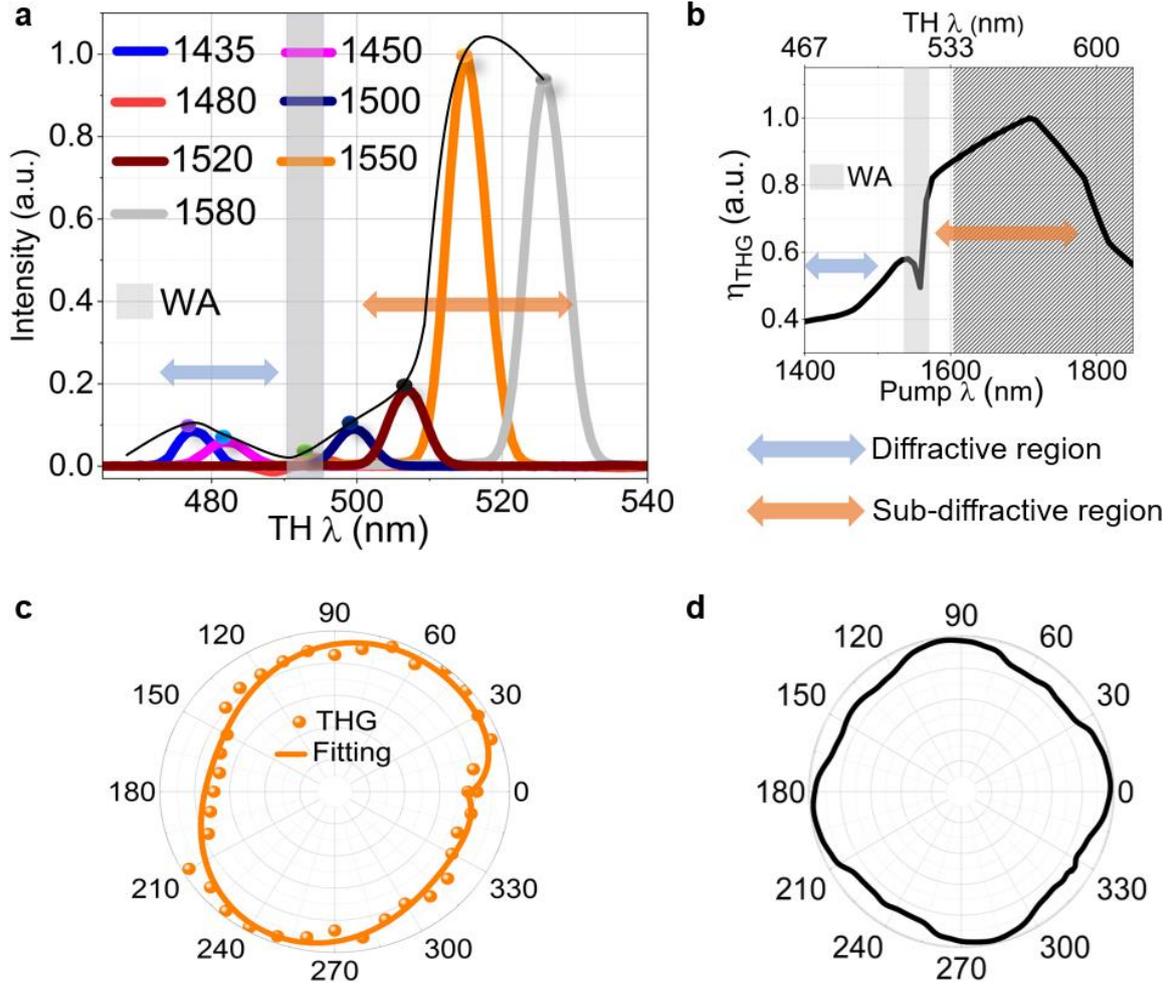

**Fig. 4. Third harmonic generation enhancement in a TMDC metasurface.** (a) Measured wavelength-dependent third harmonic response in forward direction for metasurface C. The shaded region represents the Wood's anomaly. (b) Simulated third harmonic conversion efficiency, as a function of pump wavelength, of MoS$_2$ truncated cone metasurface C. The patterned region represents the spectrum range not accessible by the laser in our experimental setup. (c) Experimental polarisation resolved THG of metasurface C, at pump wavelength of 1550 nm. (d) Simulated polarisation resolved THG of metasurface C, pump wavelength 1550 nm.

To gain a better insight on the physics of the third harmonic emission, we further perform full-wave numerical simulations of the THG in our metasurface. Figure 4b shows the simulated conversion efficiency of THG as a function of pump wavelength. The simulated THG conversion efficiency of metasurfaces A and B is presented in Fig. S4 of Supplementary Information. The numerical data, Fig. 4b, shows good qualitative agreement with the experimental data in Fig. 4a, capturing the two main effects: (*i*) There is a peak of the THG at longer wavelengths in the non-diffractive regime, at a pump wavelength of approximately 1700 nm. (*ii*) Both in the experiment and in the simulations, a drastic decrease of the TH efficiency is observed at and below the WA. We note that the higher-order Mie modes, supported by the metasurface at the third harmonic wavelengths are hindered by absorption losses. While MoS$_2$ has very high refractive index in the spectral region 450-600 nm ($n$ = 4.8-5.6), it also has a large absorption coefficient ($k$ = 3.1-1.1).

Figure 4c and 4d further show the experimentally measured and simulated polarisation resolved THG response of the metasurface C at 517 nm (pump wavelength 1550 nm). This response reveals that the THG power is nearly independent on the incident polarisation and co-linearly polarised with the pump beam. Good agreement is observed between our experiment and simulations. This observation confirms that the crystallinity of the MoS$_2$ metasurface belongs to the D$_{3h}$ symmetry group. The slight asymmetry in the experimental polarisation measurement is likely caused by a slight misalignment in the experimental setup and slight asymmetry in the fabricated samples.

Furthermore, full wave nonlinear simulations have been performed in order to investigate the tapering angle effect of the following three meta-atom shapes: (*i*) cone, (*ii*) truncated cone, (*iii*) vertical cylinder, on the THG conversion efficiency (in either direction forward and backward) and the results are presented in Fig. S5 of Supplementary Information. The THG directionality is quite similar in truncated shapes, cone and truncated conical meta-atoms. Whilst larger THG directionality is observed, in the entire spectrum (1400 nm to 1600 nm), in vertical straight cylindrical meta-atoms because of their large volume as compared to truncated meta-atom shapes.

*SHG in TMDC metasurfaces.* Next, we turn our attention on the second-order nonlinear response of the TMDC metasurfaces and study the resonant enhancement and the directionality of the SHG. The $MoS_2$ metasurface supports several resonant QNMs at the SH wavelengths, as indicated in Fig. 2. The coupling of SH light to these modes boost the SHG efficiency and promotes its radiation to the far-field[53].

To test the influence of the resonant modes on the SH enhancement, we examined the wavelength dependence of the SHG from metasurface C. For this purpose, we scan the wavelength of the infrared pump beam in the range of 1200-1600 nm. The measured SH spectra for different pump wavelengths are shown in Fig. 5a. Two peaks of SH emissions can be observed when the SH wavelength is in the vicinity of the MQ and ED QNMs. The SH enhancement, however, is significantly higher at the position of the MQ mode at 1400 nm, despite the stronger far-field scattering at the position of the ED mode. This feature of the SH enhancement can be explained by considering the near-field intensity enhancement, its overlap with the location of SH current sources and the modal overlap of the eigenmodes at the participating frequencies. The highest SH efficiency is estimated at $3.4 \times 10^{-10}$, at incident peak power of 4.3 kW, beam FWHM of 4.6 μm, see Fig. S3 of Supplementary Information.

While Ref.[41] treats the nanoresonators as bulk nonlinear media, the bulk $MoS_2$ has $D_{6h}$ symmetry and does not exhibit bulk second-order nonlinear effects. However, for a few-layer-thick material, the actual second-order nonlinear response will depend on the exact number of layers[50]. In the case of TMDC metasurfaces, the situation is even more complex due to the inhomogeneous near fields and the absorption of the $MoS_2$. Therefore, it is important to quantify the position and origin of the nonlinear sources for SHG in TMDC metasurfaces.

To determine the location of the SH current sources, we assumed four extreme scenarios on how the second order nonlinearity ($\chi^{(2)}$) is present in TMDC metasurfaces. These include having nonlinear sources: (*i*) at top and bottom surfaces, (*ii*) in the bulk, (*iii*) at the top surface only and (*iv*) at the bottom surface only. To characterise these four cases, we performed full-wave nonlinear simulations of the metasurface SHG response. The analysis of these four scenarios is discussed in detail in the Supplementary Information (see Fig. S6). Among these four scenarios, the first one best describes our experimental SH response. Namely, the SH response is a result of the interference of the SHG from the *top and bottom layers* of the $MoS_2$, as depicted in Fig. 5b. This is in stark contrast to the surface SHG emission from nanostructures made of centrosymmetric materials, such as silicon metasurfaces[54]. In such nanostructures the second harmonic emission originates mainly from their side walls, where the symmetry is broken. In our TMDC metasurface, the dominant contribution is from the top and bottom $MoS_2$ layers, which are approximately four orders of magnitude more efficient.

The corresponding simulations for the wavelength dependence of the SH efficiency from metasurface C are presented in Fig. 5c. A good agreement can be observed with the measured SHG (Fig. 5a). We note that our theoretical calculations show the strongest SHG at the position of the MD mode, which are not accessible by our tunable pump laser. To quantify this wavelength dependence, we further performed coupled mode analysis on the nonlinear wave interactions. The SHG efficiency depends upon the coupling between the SH light and eigenmodes. It is well known that in nanostructured resonant systems, the SHG efficiency, which is the ratio of SH to pump power, depends on how large is the spatial field overlap between the pump and SH field in the regions occupied by nonlinear sources. Indeed, using the reciprocity theorem, a simple expression of the SHG efficiency can be obtained in terms of overlap integrals[8]

$$\eta_{SHG} \sim \left| \int_S \boldsymbol{J}_{NL}(2\omega) \cdot \boldsymbol{E}_{SH} dS \right|, \quad (3)$$

where $\boldsymbol{J}_{NL}(2\omega) = -i\omega_{SH}\epsilon_0 \chi^{(2)}_{MoS_2}[(E_{p,x}^2 - E_{p,y}^2)\hat{x} - 2E_{p,x}E_{p,y}\hat{y}]$ is the nonlinear surface current source induced by the pump at the SH wavelength. The aforementioned expression of the $\boldsymbol{J}_{NL}(2\omega)$, that we consider in the model and in nonlinear simulations stems from the tensor nature of $MoS_2$, which has the following nonzero elements: $\chi^{(2)}_S(MoS_2) = \chi^{(2)}_{xxx} = -\chi^{(2)}_{xyy} = -\chi^{(2)}_{yxy} = -\chi^{(2)}_{yyx}$. $\boldsymbol{E}_{SH}$ is the field induced on the surface of the resonator by a far-field virtual source, i.e., a plane wave located at the detector's position. The integral is performed over the top and bottom surfaces of the $MoS_2$ meta-atoms. The pump field, $\boldsymbol{E}_p$, and the SH field, $\boldsymbol{E}_{SH}$, can be retrieved by solving the linear problems with a plane-wave excitation at the pump wavelength. Since in our problem the modes are resonantly excited at the SH wavelength, we only expand the SH field in QNMs, and we retain only the three QNMs that are present in the SH spectrum under investigation, i.e., the MD, the ED and the MQ-like modes.

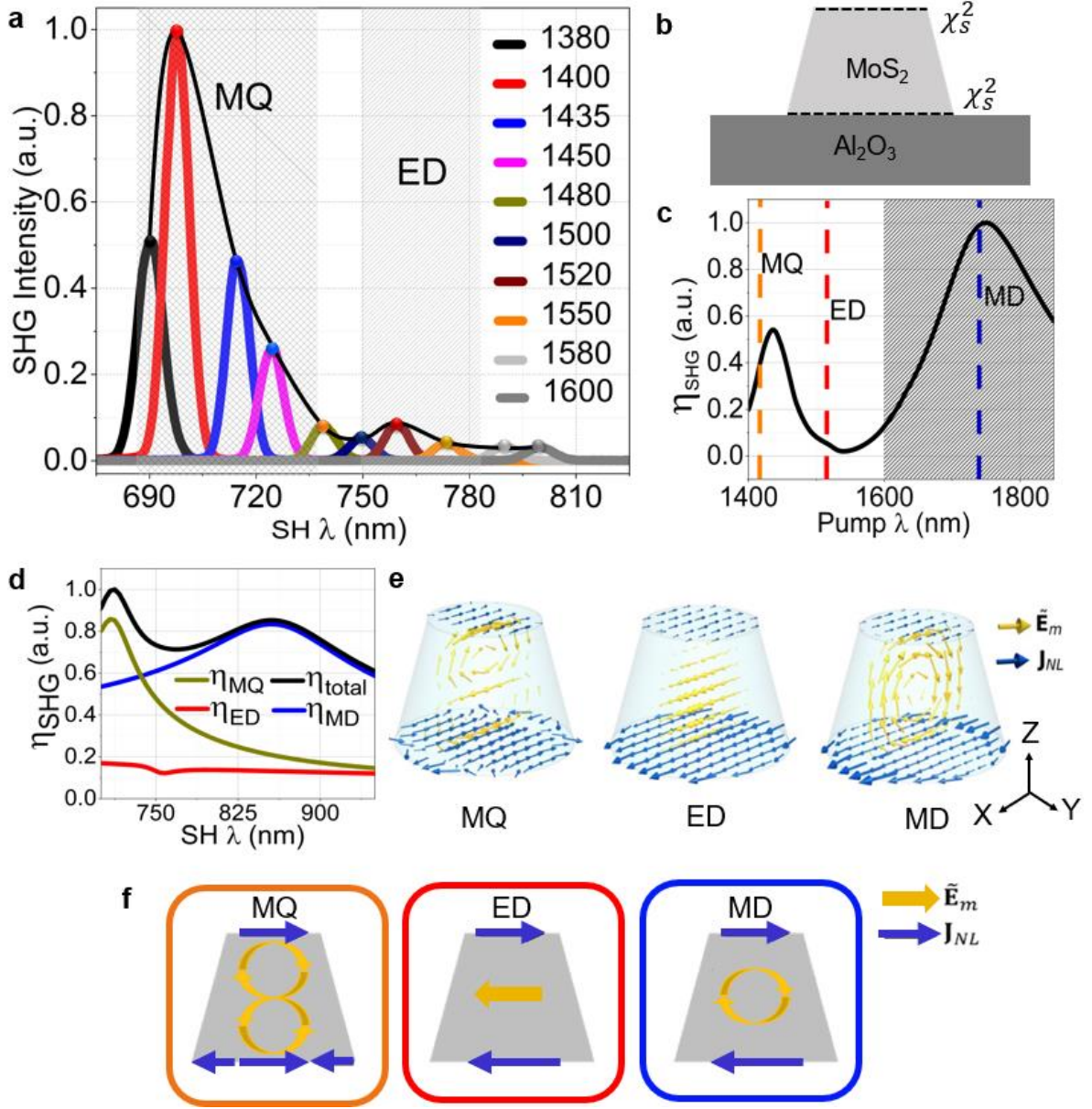

**Fig. 5. Second harmonic generation in a TMDC metasurface.** (a) Experimentally measured wavelength dependence of the SHG from metasurface C (b) Schematic showing the locations of the SH current source ($\chi^2$). (c) Calculated SHG conversion efficiency by employing full-wave nonlinear simulations. The patterned region is the spectral range not accessible by the laser in our experiments. (d) SHG conversion efficiency calculated using quasi-normal modes couple-mode theory. (e) Spatial overlap between the induced nonlinear current source ($J_{NL}$) and eigenfields of the calculated quasi-normal modes ($\widetilde{E}_m$). (f) Schematic of the modal overlap of nonlinear current source and eigenfields.

The metasurface response at the pump wavelengths (1200-1600 nm) is non-resonant, and we can therefore retain the full-wave expression of $J_{NL}(2\omega)$ without expanding the pump field $E_p$. As a result, the expression of the SHG efficiency can be recast as follows:

$$\eta_{SHG} \sim \left| \alpha_{MD} \int_S J_{NL}(2\omega) \cdot \widetilde{E}_{MD} dS + \alpha_{ED} \int_S J_{NL}(2\omega) \cdot \widetilde{E}_{ED} dS + \alpha_{MQ} \int_S J_{NL}(2\omega) \cdot \widetilde{E}_{MQ} dS \right|. \quad (4)$$

The expression in Eq. (4) is only based on the linear properties of the metasurface and is limited to only three QNMs. As such, it does not give an exact quantitative description of the SHG efficiency, however, it provides useful insights into the origin of SH light in our system. In particular, Eq. (4) predicts the presence and the nature of SHG spectral peaks and allows to quantify the coupling strength of SH light to each QNM. Furthermore, Eq. (4) can be used to optimise the meta-atom shape and size in order to enhance or inhibit the response of specific QNMs or to favour the interaction between different QNMs (see Fig. S7 of Supplementary Information).

In Fig. 5d we also show the calculated SHG conversion efficiency using the coupled QNM theory, Eq. (4), as well as the contributions of each of the three QNMs to the overall efficiency,

$$\eta_m = \left| \alpha_m \int_S \mathbf{J}_{NL}(2\omega) \cdot \tilde{\mathbf{E}}_m dS \right|, \qquad (5)$$

where $m = MD, ED, MQ$. The calculation is performed by considering a pump field, linearly polarised along the *x*-axis and the co-polarised component of the SHG. We can clearly observe that the SHG displays peaks of efficiency that are associated only with the two magnetic modes, while SH light coupling with ED-like mode is significantly inhibited. This result is in qualitative agreement with the full-wave nonlinear numerical simulations (Fig. 5c) and our experimental observations (Fig. 5a).

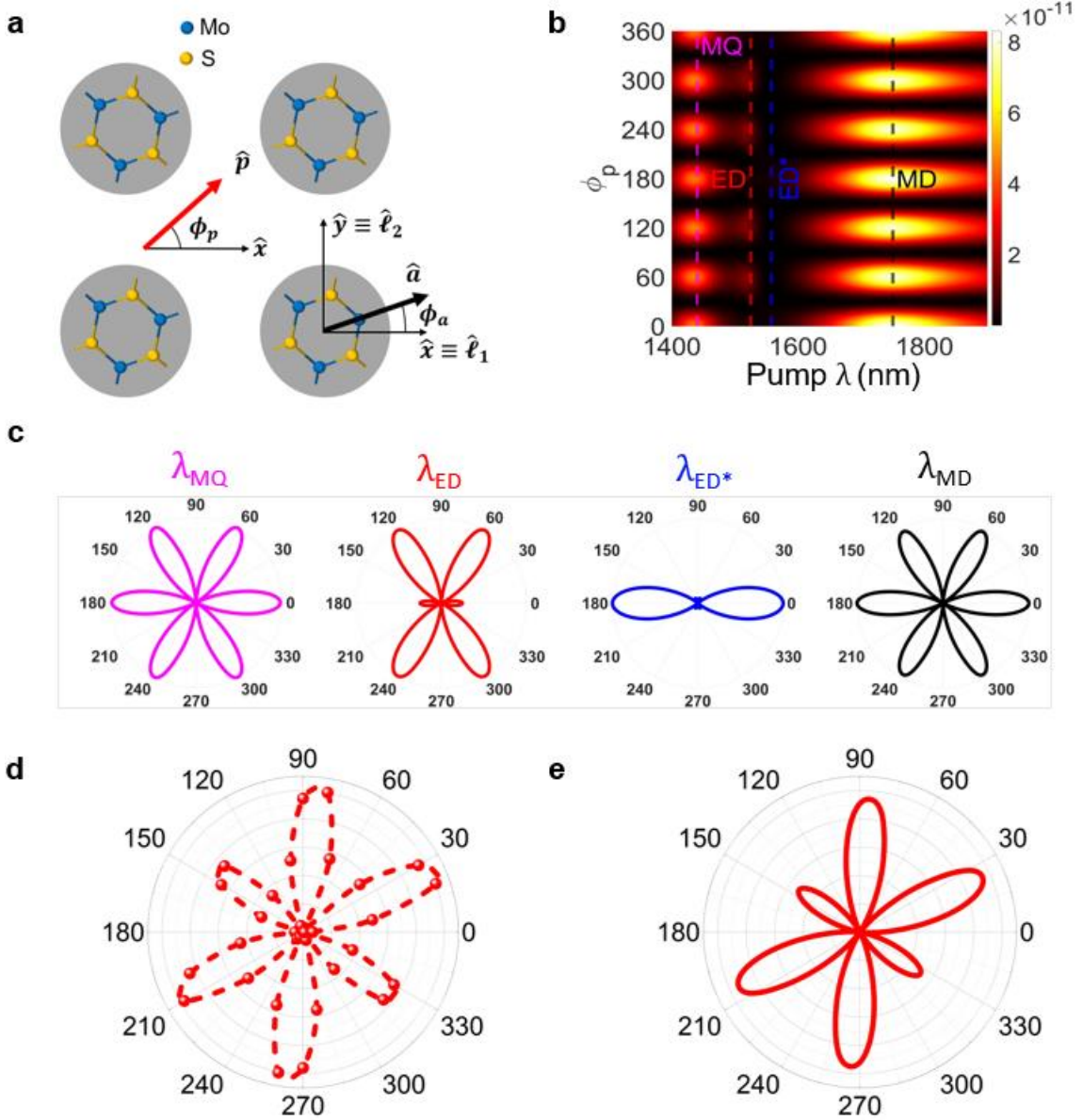

**Fig. 6**. **Polarisation-resolved SHG in a TMDC metasurface.** (a) Schematic of the crystal orientation of the MoS$_2$ with respect to the metasurface lattice and incident polarisation. (b) Simulated polarisation resolved SHG, in the forward direction, as a function of $\phi_p$ and pump wavelength when $\phi_a = 0$ (c) Polarisation resolved SHG as function of $\phi_p$ (when $\phi_a = 0$), at the wavelengths marked MQ, ED, ED* and MD, where ED* is the point in close vicinity of ED, as shown in (b). (d) Measured polarisation resolved SHG as a function of $\phi_p$ (when $\phi_a = 30°$) of metasurface C at ED. (e) Simulated polarisation resolved SHG as function of $\phi_p$ (when $\phi_a = 30°$) of metasurface C at ED.

The reason behind the dominant role of magnetic modes over the electric mode is better clarified by inspecting the overlap integrals that appear in the expressions of $\eta_m$. In Fig. 5e, we show the spatial distribution of the eigenmodes $\tilde{\mathbf{E}}_m$ along with the induced nonlinear surface currents $\mathbf{J}_{NL}$. When the SH is tuned to the wavelengths of the magnetic QNMs, $\tilde{\lambda}_{MD}$ and $\tilde{\lambda}_{MQ}$, the spatial overlap between the nonlinear source and the eigenfield is optimal on both the top and bottom surface (see MQ and MD in Fig. 5e). On the contrary, this overlap is very weak for the ED mode (see ED in Fig. 5e). Although the ED mode displays strong field localisation at the centre of the meta-atom's volume, the eigenfield is less intense at the top and bottom surfaces, where the nonlinearity is present. In addition, the SH coupling to the ED mode is frustrated by the fact that the nonlinear current source $\mathbf{J}_{NL}$ at $\tilde{\lambda}_{ED}$ has odd parity, while the eigenfield $\widetilde{\mathbf{E}}_{ED}$ has even parity. This difference in the parity of the mode leads to a tiny overlap integral, as shown in the schematic of Fig. 5f.

An important consideration for the SHG from TMDC metasurfaces is the mutual arrangement of the MoS$_2$ crystalline orientation with respect to the metasurface lattice and the incident polarisation. This arrangement is depicted in Fig. 6a, where the lab-frame is oriented along the metasurface lattice, $\hat{\mathbf{p}}$ is the pump polarisation unit vector, $\hat{\mathbf{a}}$ is the armchair direction unit vector, $(\hat{\boldsymbol{\ell}}_1, \hat{\boldsymbol{\ell}}_2)$ are the primitive vectors of the metasurface lattice, $(\hat{\mathbf{x}}, \hat{\mathbf{y}})$ are the lab-frame axes, $\phi_a$ is the angle between the armchair direction and the lab-frame, and $\phi_p$ is the angle between the pump polarisation and lab-frame. The SHG efficiency from the TMDC metasurface, $\eta_{SHG}$, is a function of both $\phi_p$ and $\phi_a$. In the numerical simulations, we fix $\phi_a = 0$ and evaluate the SHG conversion efficiency as a function of the pump wavelength and the angle of polarisation $\phi_p$ in the forward (Fig. 6) and backward (see Fig. S8 of Supplementary Information) directions.

This polarisation/wavelength-resolved analysis confirms the conclusion that when the SH wavelength is either at MQ or MD QNMs (magenta and black dashed vertical lines in Fig. 6b, respectively), the SHG efficiency is maximised. In this situation, there are six peaks of equal strength when the incident polarisation is varied. These peaks correspond to polarizations perpendicular to the armchair direction of the MoS$_2$ (magenta and black plots, labelled $\lambda_{MQ}$ and $\lambda_{MD}$ respectively in Fig. 6c), meaning that there is only one radiating SH mode because of an optimal overlap of SH light with the eigenfields of magnetic modes. On the other hand, when the SH wavelength is near the ED or ED* wavelengths (red and blue dashed vertical lines in Fig. 6b), the ED mode is weakly coupled to the SH source and therefore, the SH originates from the interference of the three QNMs, rather than only the ED mode. In fact, depending on the exact wavelength near the ED, one can have stronger dominance of the ED mode (red plot, labelled $\lambda_{ED}$ in Fig. 6c, with four stronger peaks and two weaker peaks) or suppression of the ED mode (blue plot, labelled $\lambda_{ED^*}$ in Fig. 6c, with two stronger peaks and four weaker peaks). It is worth mentioning that the coupling of the SH light to the electric and magnetic modes depends upon the shape, location of nonlinear current source and symmetry of the meta-atoms, which eventually modulate the shape of the polarisation resolved SHG. As such, the SH emission dramatically depends on the tapering angle of the meta-atoms, which in the experiment is fixed by the fabrication. The simulated SHG in meta-atoms of two other tapering angles, e.g. conical meta-atoms and straight cylindrical meta-atoms are depicted for completeness in Fig. S9 of the Supplementary Information.

To verify the polarisation dependence of the SHG from TMDC metasurfaces, we have also performed experiments on polarisation resolved SHG in forward direction at the pump wavelength of 1550 nm. The measured polarisation resolved SHG is shown in Fig. 6d. It demonstrates six-fold rotational symmetry, when $\phi_p$ varies from 0 to 360 degrees. There is a very good agreement between our experimental (Fig. 6d) and simulated data (Fig. 6b and 6c). A comparison of the simulated and experimental results reveals that, in the experiment, the actual armchair direction is at $\phi_a \approx -30°$. Indeed, when in simulations we set $\phi_a = -30°$, we obtain excellent agreement with the experimental polarisation-resolved SHG, as shown in Fig. 6d and 6e. For the chosen pump wavelength generating SH in the vicinity of the ED mode, instead of six peaks of equal amplitude, in Fig. 6d and 6e we can observe four stronger peaks.

Finally, the most intriguing feature of the presented MoS$_2$ metasurface is the capability to tune the directionality of the SH emissions and switch the emission between forward and backward direction. While SH emission has been tested for various dielectric metasurfaces[16,17,23,24,55,56], no directionality-switching capability has been reported so far. In Ref.[31] switching of the SH emission has been observed for a single (110) GaAs nanoantenna. However, no metasurfaces of such nanoantennas have yet been realised to date. Furthermore, all metasurfaces to date that are resonant at the fundamental wave have multiple diffraction orders at the SH wave, which would not allow for pure forward to backward switching of the SH emissions. Here, for the first time to our knowledge, we demonstrate a dielectric metasurface with forward to backward continuous tunability of the SH emissions. This property is a direct consequence of the position of the nonlinear sources and the interference of the QNMs in the far-field. Importantly, to tune the SH directionality, no physical change in the structure or complicated geometry is required. The tuning can be achieved by varying the input wavelength or polarisation.

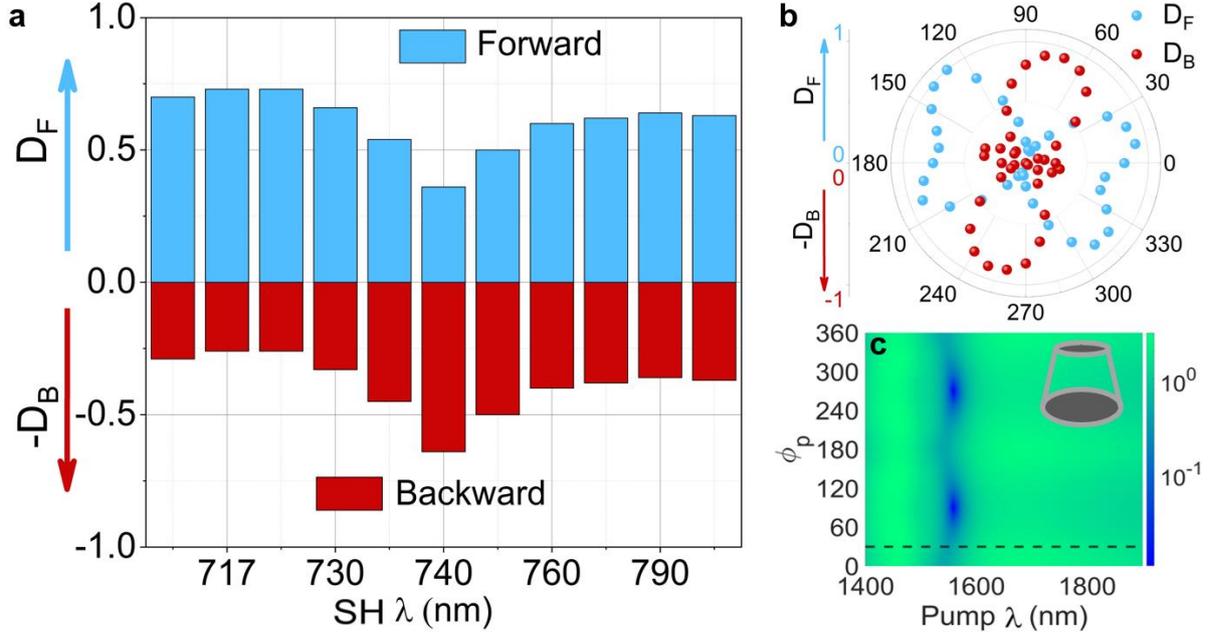

**Fig. 7. Directionality switching of the second harmonic emission.** (a) Measured SH forward and backward directionality of emission as a function of the second harmonic wavelength, when $\phi_p = 0°$. (b) Measured forward and backward SH directionality at switching wavelength, 1480 nm, as a function of $\phi_p$ when $\phi_a = -26°$. (c) Simulated directionality mapping of forward and backward SH emissions from the truncated cone (as shown in the inset) metasurface, as a function of pump wavelength and $\phi_p$, when $\phi_a = 0°$. (c) Simulated SH directionality ($P_F^{SH}/P_B^{SH}$), as ratio of forward and backward SH emissions, of truncated cone meta-atom as a function of pump polarisation at point of switching wavelength.

To quantify such tuning of the SH emission, we define the SH directionality in terms of the relative power of the SH emitted in forward (F) direction as $D_F = \frac{P_F^{SH}}{P_F^{SH}+P_B^{SH}}$ and in backward (B) direction as $D_B = \frac{P_B^{SH}}{P_F^{SH}+P_B^{SH}}$, where $P_F^{SH}, P_B^{SH}$ are the measured normalized SH powers in forward and backward direction, respectively. In Fig. 7a, we show the experimentally measured SH directionalities, $D_F$ in blue and $D_B$ in red, as a function of the pump wavelength, for $\phi_p = 0$ and $\phi_a = -30°$. In the bar charts, 0.5 is the threshold value, which defines the switching of SH emissions from forward to backward direction when $D_F < 0.5$ and $D_B > 0.5$. In Fig. 7a, the SH directionality switching point can be observed around 740 nm. In Fig. 7a, the *x*-axis scaling (representing SH wavelengths) corresponds to the selected fundamental wavelengths, where our OPO shows stable performance in the spectral range from 1400 nm to 1600 nm. Figure 7b depicts the measured SH directionality, $D_F$ and $D_B$, at switching point as a function of $\phi_p$. We further complement our measurements by numerical simulations and in Fig. 7c we show the simulated SH directionality mapping, $(P_F^{SH}/P_B^{SH})$, as function of pump wavelength and $\phi_p$ for $\phi_a = 0$. The measured data in Fig. 7a corresponds to the horizontal dashed line in Fig. 7c. The switching point of the SH directionality as a function of $\phi_p$ is observed for two values $\phi_p \cong 80°$ and $\cong 260°$. In Fig. 7c it can also be seen that the SH directionality switching occurs in the close vicinity of the ED mode, which can be triggered either as a function of the pump wavelength or polarisation. In contrast, the directionality of the SH emission is symmetric at the MQ and MD resonances. A good agreement can be observed between the experimental and simulated directionality data. The difference between the measured and simulated data might be because of the directionality being sensitive to the exact shape of the fabricated $MoS_2$ meta-atoms, such as the tapering angle and the surface roughness.

To illustrate, the effect of the tapering on the switching of the SH directionality in Fig. 8 we have simulated the SH directionality (see also Fig. S10 of Supplementary Information) when the $MoS_2$ meta-atom has the shape of a straight vertical cylinder (no sidewall tapering). The data in Fig. 8a corresponds to the horizontal dash line in Fig. 8c. It may be seen that in this cylindrical geometry, one can achieve a nearly complete switching of the SH emission from forward (~725 nm) to backward (~825 nm) direction, as shown in Fig. 8a. Such nonlinear switching is useful for application in nonlinear mirrors and laser mode-locking[34]. Additionally, in case of vertical straight cylindrical metasurface, the SH directionality as function of $\phi_p$ shows no switching at the wavelength of 825 nm. This polarisation dependence of the directionality is shown in Fig. 8b and corresponds to the vertical dash-dotted line in Fig. 8c.

Furthermore, in Fig. 8d-f we show the effect of the tapering angle on the second harmonic directionality. Three shapes with three different tapering angles of the meta-atoms have been calculated and compared in Fig. 8d-f. It is seen that the contrast of the unidirectional emission can be controlled with the tapering angle of the meta-atoms. However, the tapering angle also affects the switching-point wavelength and the spectral bandwidth of the switching. These changes are likely due to the spectral shifts of the resonances. In our experimental case, the switching ratio is lower that in the other two cases, however it is also spectrally broader and bell-shaped, which has improved the robustness of this condition to experimental imperfections.

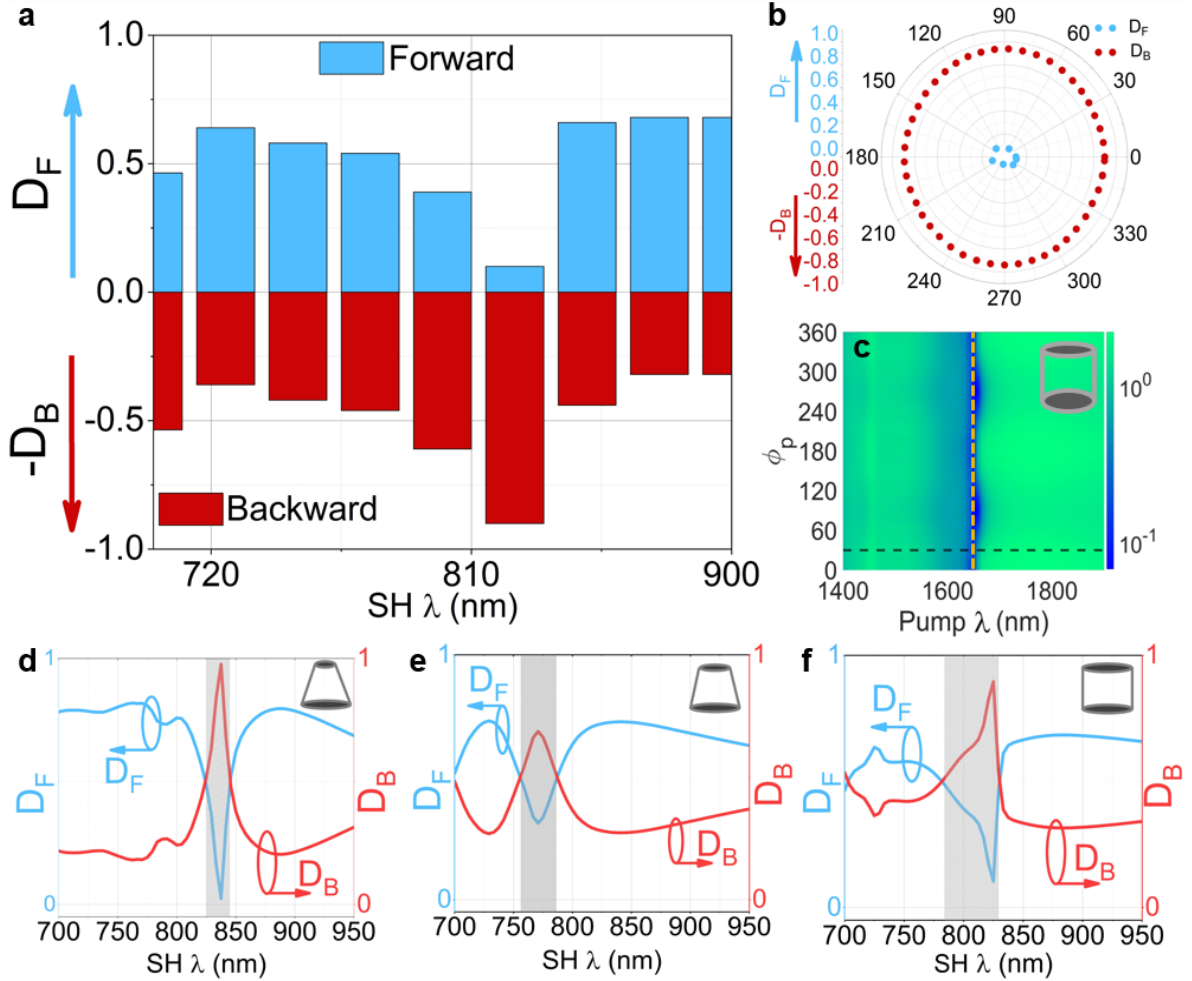

**Fig. 8**. **Tapering angle effect on directionality switching of the second harmonic emission.** (a) Simulated SH forward and backward ratio of MoS$_2$ metasurface with straight vertical walls of cylindrical meta-atoms, as a function of SH wavelength. (b) Simulated forward and backward SH directionality of vertical straight cylindrical metasurface as a function of $\phi_p$ at a pump wavelength of $\cong 1650$ nm. (c) Simulated directionality ($P_F^{SH}/P_B^{SH}$) mapping, as ratio of forward and backward SH emissions, of a straight cylindrical (as shown in the inset) MoS$_2$ metasurface, as function of pump wavelength and $\phi_p$, when $\phi_a = 0°$. (d-f) Comparison of calculated ratio of SH emissions in forward and backward direction as function of pump wavelength. (d) Cone meta-atom (e) Truncated cone meta-atom (f) Vertical cylinder. The shaded area shows the switching point of the directionality.

**Discussions**

In summary, we have demonstrated, for the first time to our knowledge, the directional nonlinear harmonic generation in a high-refractive index single-crystal vdW MoS$_2$. The vdW interactions between the TMDC layers offer unique opportunity to realise such metasurfaces on any smooth and transparent substrate, thus one can excite and detect the SH and TH signals from each side of the metasurface. We have experimentally and theoretically demonstrated that by changing the aspect ratio of the truncated-cone meta-atoms, one can tune the induced resonances of the metasurface and provide direct control over the second- and third-order nonlinear generation. As compared to other dielectric and semiconductor metasurfaces the use of the high-index TMDCs allows to design highly subwavelength resonators and therefore to construct nonlinear metasurfaces with single-beam emission at the harmonic waves. In this sub-diffractive regime only a zero[th] diffraction order is emitted from the metasurface, which makes it act as a homogeneous ultra-thin nonlinear crystal but with enhanced conversion efficiency. Most importantly, the second-harmonic emission can be tuned from forward to backward direction, controlled by wavelength, incident polarisation and metasurface geometry. This tunable unidirectional nonlinear emission facilitates novel applications in nonlinear light sources.

Additionally, we have employed Mie-resonances at the harmonic wavelengths as a versatile tool to control both SHG and THG at the nanoscale. The choice of resonances at the harmonic wavelengths further makes our metasurfaces deeply subwavelength for both the second and third harmonic waves. Our work aims to demonstrate the fundamental physical phenomena on the nonlinear emission in TMDC metasurfaces and as such is able to unveil the origin of the quadratic nonlinearities in TMDC metasurfaces. These stem from the broken crystalline symmetry at the top and bottom layer of the meta-atoms. This feature is distinct to the harmonic generation by surface nonlinearities in other bulk non-centrosymmetric material, such as silicon and results in more than two-orders of magnitude higher SHG efficiency with respect to silicon metasurfaces. We note that while our metasurface design does not aim at reaching record-breaking conversion efficiencies, it is possible to applied multiple strategies for increasing the efficiency of both SHG and THG. These include designs with higher quality resonances, such as in the case of bound-state in the continuum[57].

The comparison of the performance of our TMDC metasurface with other existing works, in terms of its high index and SH directionality switching, has been presented in the table I. The cross symbol in table I refers to the lack of SHG/THG measurement or no SHG/THG in those references. We note that while the nonlinear emission from monolayer TMDCs coupled to plasmonic[58,59] or dielectric metasurfaces[60] have also been recently studied, our TMDC metasurface offers several fundamental advantages. These include the ability to simultaneously enhance the SHG and THG processes, offering a significantly higher interaction volume for the THG. For the case of SHG, it further offers the ability to switch the directionality of SH emission between forward and backward direction, which so far has not been demonstrated in other metasurface systems. Importantly, this intriguing feature has been achieved all-optically: there is no need to apply physical changes to the metasurface and can be dynamically controlled by varying the pump wavelength or polarisation. We have experimentally proven this SH directionality switching and have theoretically shown that the contrast of this switching can be further enhanced for the case of vertical cylinders, thereby virtually acting as mirror for the SH light.

TABLE I: *Performance comparison of the $MoS_2$ metasurface with metasurfaces of other high index dielectrics.*

| References | Nonlinear Media | Refractive index (n) (700 nm – 800 nm) | Extinction coefficient (k) (700 nm – 800 nm) | SHG | THG | SH directionality | Periodicity | Fundamental wavelength |
|---|---|---|---|---|---|---|---|---|
| Ref.[55,61,62] | LiNbO$_3$ | 2.2 | 0 | ✓ | ✓[55] ✗[61,62] | ✗ | 590 nm 900 nm[61] 590 nm[62] | 815 nm 1550 nm[61] 830 nm[62] |
| Ref.[16-17] | Si | 3.7-3.6 | 0.01 | ✗ | ✓ | ✗ | 750 nm[14] 1280 nm[15] | 1350 nm[14] 2320 nm[15] |
| Ref.[24] | AlGaAs | 3.7-3.6 | 0.14-0.09 | ✓ | ✓ | ✗ | 840 nm | 1570 nm |
| Ref.[23] | GaAs | 3.7 | 0.15-0.09 | ✓ | ✗ | ✗ | 600 nm | 1020 nm |
| **This work** | **MoS$_2$** | **5.6-4.7** | **0.4-0** | **✓** | **✓** | **✓** | **300 nm** | **1550 nm** |

**Methods**

*Device Fabrication:* We fabricated multiple arrays of MoS$_2$ nanoresonators (MoS$_2$ metasurfaces) on a sapphire (Al$_2$O$_3$) substrate. To achieve pristine quality of the sample, we mechanically exfoliated vdW material MoS$_2$ onto the substrate. The thickness of the transferred flakes was then measured by surface profilometer. After cleaning, we spin coated a single layer of the ZEP positive resist followed by a thin layer of conductive polymer (to dissipate the charges during the lithography process) on the selected flake. The flake was subsequently patterned into nanoresonators using electron-beam lithography (EBL). After developing the sample, we deposit 80 nm thick layer of the Al (for dry etching mask), using e-beam evaporator. The truncated cone nano-pillars were then dry etched by employing inductive-coupled plasma of SF$_6$+CHF$_3$+Ar. and then Al caps from the top of the truncated-cone pillars removed by lift-off process. Finally, we employed scanning electron microscopy (SEM) to examine the quality and dimensions of the MoS$_2$ metasurfaces.

*Numerical Calculations:* To theoretically describe our experimental results, first, we performed FDTD simulations using Lumerical FDTD tool for the manifestation of mode properties in the SH spectrum (700-900 nm). We used periodic boundary conditions along with normal-incidence plane-wave illumination from either side (the air or substrate side) of the metasurface and calculated the linear spectra of the metasurface from both sides. The data presented in the paper corresponds to the plane-wave illumination from the air side to match with our experimental arrangement. We used the refractive index and extinction coefficient of the $MoS_2$ bulk from a work presented in Ref.[63] (see Supplementary Information Fig. S1). For the nonlinear simulations we modelled the truncated cone meta-atoms in COMSOL Multiphysics. We employed full-wave nonlinear simulations in COMSOL (finite-element method) to extract the SH and TH efficiency in the range of pump wavelength. In simulations, the pump is incident from the air side (as in the experiment), *s*-polarised in the *x*-direction ($\emptyset_p = 0$). Moreover, we use pump intensity of 1 $GW\,cm^{-2}$ and $\chi^2$ ($MoS_2$) tensor value of 3 $nm^2V^{-1}$ in agreement with experimental measurements presented in Ref.[64]. To gain deeper insight of the physics of SH peaks, we employed QNMs theory and solved the complex eigenvalue problem with a finite-element method (COMSOL).

**Acknowledgements:**
This work was supported by The Punjab Educational Endowment Fund (PEEF) Pakistan, by the Australian Research Council through Centres of Excellence (CE20010001), Discovery Projects (DP190101559) and ARC Early Career Researcher Award (DE190100430) programs, by The Royal Society and the Wolfson Foundation. We acknowledge the use of the ACT node of the Australian National Fabrication Facility (ANFF).